\documentclass[12pt]{article}
\usepackage{amsfonts}
\usepackage{amsmath, graphicx, epsfig,psfrag, verbatim}
\usepackage{eucal,  amssymb,amsmath,amsthm,graphicx, amsfonts, latexsym}
\usepackage[utf8]{inputenc}

\newtheorem{result}{Result}[section]

\newtheorem{defin}{Definition}[section]

\begin{document} \parskip=5pt plus1pt minus1pt \parindent=0pt
\title{Epidemic models on social networks\\ -- with inference}
\author{Tom Britton$^1$, Stockholm University}
\date{\today}
\maketitle

\begin{abstract}
Consider stochastic models for the spread of an infection in a structured community, where this structured community is itself described by a random network model. Some common network models and transmission models are defined and large population proporties of them are presented. Focus is then shifted to statistical methodology: what can be estimated and how, depending on the underlying network, transmission model and the available data? This survey paper discusses several different scenarios, also giving references to publications where more details can be found.

\end{abstract}


\footnotetext[1]{Department of Mathematics, Stockholm University, 106 91 Stockholm, Sweden.\\ Email: tom.britton@math.su.se}

\section{Introduction}\label{sec-intro}

In the current paper we are concerned with stochastic models for how an infectious diseases spreads in a community, where the social structure of relevance for the disease spreading is described by a random network model. 

In certain cases the underlying social structure is known, the presence of households being the prime example. Here we put more focus on the case where the underlying structure is not entirely known, which explains why a random network model is advocated.

Which network model to use will depend on the infectious disease under consideration and the community upon which it spreads. If considering diseases with airborne spreading like influenza and childhood diseases, the network should reflect pairs of individuals being in proximity of each other on regular basis (preferably also adding random contacts). If spreading occurs through close physical contact such as Ebola, the network edges connect pairs of individuals having such contacts on regular basis, and if considering a sexually transmitted infection (STI) the underlying network will be that of sexual contacts.

The type of network model to be used hence depends on the disease and context. If we are considering short term outbreaks a static network may be sufficient, whereas if we are interested in longer time spans a dynamic model, where individuals die and new are born but also where connections are dropped and new are created, may be preferred. Individuals are usually not all the same: some individuals have more social contacts than others, which should be reflected in the \emph{degree distribution} (the degree of an individual is the number of social connections it has). Other characteristics of a network is \emph{clustering}, reflecting how common triangles in a network are, and degree correlation: positive (negative) degree correlation implies that individuals with high degree tend to mix with individuals of high degree. Given a set of such \emph{egocentric} network proporties, the network is then often defined by saying that, other than obeying the pre-specified proporties the network is chosen randomly among all networks satisfying the properties. Of course, more complicated networks may also be considered, for example by allowing different types of individuals, and having weighted edges affecting the transmission probabilities.

Our interest lies in how an infectious model can spread on the random network. There are different infectious disease models, the base model being an SIR (Susceptible-Infectious-Recovered) epidemic model, where individuals are first susceptible, and if they get infected they become infectious and later they recover and stay immune for the rest of the study period. Other version are SEIR, where exposed individuals are first latent before they become infectious, SIRS where immunity eventually vanishes and the individual becomes susceptible again, and so on. In reality, infectivity usually builds up over time and after some further time starts dropping down to 0. Here we simplify the situation by assuming that an infected individual has constant infectivity during the whole infectious period $I$, and we focus on the two situations where $I$ is either fixed and the same for all, or otherwise exponentially distributed. The two models are referred to as the Reed-Frost and the Markovian versions. The Reed-Frost model is often studied in its discrete time version where infections happen sequentially in generations. The Markovian model assumes that infectious individuals infects each of their susceptible neighbour independently at rate $\beta$ and recovers at rate $\gamma$, and the Reed-Frost model assumes that an individual infected in generation $k$ infects each if its susceptible neighbours in the next generation independently with probability $p$ and then recovers. In both models it is possible to also allow for transmission with randomly chosen individuals beside the neighbours in the network.

One reason for studying epidemic models on networks is to better understand what model features affect spreading the most, and in particular how it is possible to reduce spreading by means of public health  measures such as vaccination, (quicker) diagnosis and treatment, isolation, travel restrictions and so on. This can be achieved by inserting the relevant preventive measure into the model, and to analyse the outcome and then compare with the outcome without prevention. 

In order to make conclusion about real life situations it is necessary to fit the models to the real world situation, preferably by collecting network and/or disease data to infer model parameters using proper statistical methods. If the entire underlying network is observed, this is often straightforward. However, as mentioned above this situation is not a common situation. Instead some egocentric data may be available, perhaps together with some outbreak data, from which to perform inference, and then statistical methods are more complicated and many problems remain open.

In the current paper we will describe such network models, transmission models "on" the networks, models capturing control measures, and their inference procedures. Needless to say, this whole area is bigger than can be captured in one review paper, so we will only touch upon most models and methods and leave out several. Another focus of the paper is to describe important unsolved problems, with the aim to stimulate more work in this important research area.

We start by describing a few different random network models (Section \ref{Net-models}), then describe the two transmission models mentioned above in a bit more detail, followed by models for prevention (Sections \ref{Inf-models} and \ref{Prevention}). In Section \ref{Properties} we present known properties of the models with and without control measures, and in Section \ref{Inference} we describe how to perform inference for several different models and data settings, also mentioning several open problems.

\section{Social network models}\label{Net-models}

\begin{figure}[ht]
\centering
\includegraphics[scale=0.6]{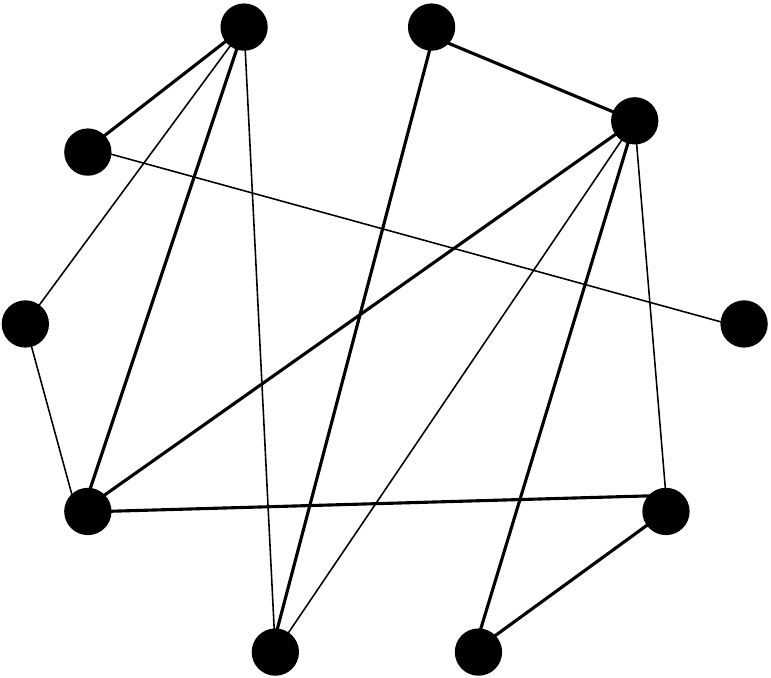}
\caption{Illustration of a small random social network. In this network nodes have degrees between 1 and 4, and the mean degree equals $E(D)=\mu=2.7$.}\label{fig_network}
\end{figure}

A network consists of nodes and edges. In our application the nodes will be individuals and the edges, between pairs of individuals, reflect some type of social relationship. Unless otherwise mentioned, we consider static, unweighted, undirected edges, all being of the same type as shown in Figure  \ref{fig_network}. In what follows we will assume that there are $n$ nodes and that which pairs of nodes that are connected by an adge is random. The focus lies on the population size $n$ being large and that the number of edges connecting pairs of individuals is of the same order $O(n)$ as the number of nodes, implying that each individual has a mean degree $E(D)=\mu$ ($0<\mu <\infty$) which remains fixed as $n\to\infty$, denoted sparse graphs/networks.

There exists several well-known random network models (the synonym ''random graph models'' is often used in more mathematically oriented papers). 

The first and most well-studied random network model is the \textbf{Erd\H{o}s-R\'{e}nyi random graph} (Erd\H{o}s and R\'{e}nyi, 1959). This model has the least possible structure. It assumes that every pair of individuals is connected to each other, independently, with probaility $\lambda /n$. This model contains a single parameter $\lambda$ being (approximately) the mean degree of individuals. An individual has $n-1$ possibly connections, each being present with probability $\lambda/n$, so the number of neighbours any individuals has is binomially distributed: $D\sim Bin (n-1, \lambda/n)$. As $n\to\infty$ it is well-known that this distribution tends to the Poisson distribution with mean $\lambda$, so $D\approx Po(\lambda)$. 

\begin{figure*}[ht]
\centering
\includegraphics[scale=0.6]{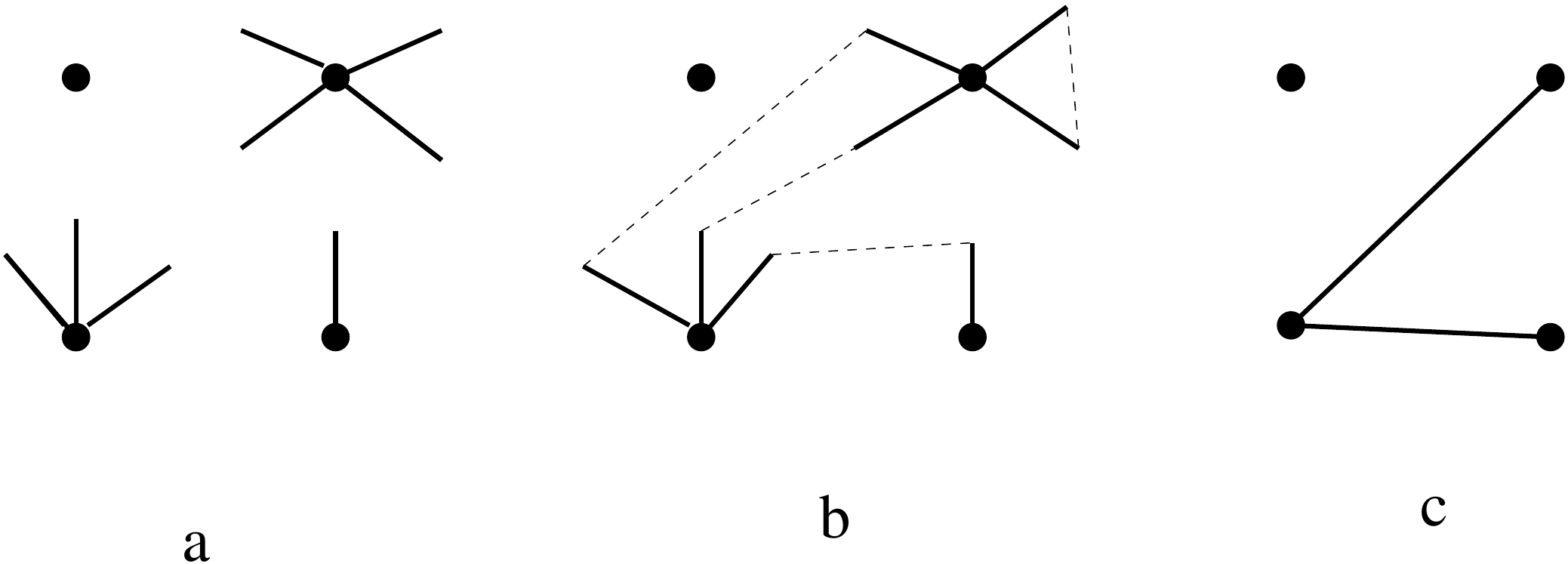}
\caption{Illustration of the configuration model for a very small network. a) The degree of each vertex is drawn i.i.d.\ from the degree distribution. b) Stubs are paired completely at random. c) Multiple edges and self-loops are removed thus producing the final network.}\label{fig_config-model}
\end{figure*}

A second well-studied model is the \textbf{Configuration model} (e.g.\ Molloy and Reed, 1998, or Bollob\'{a}s, 2001). This model is specified by an arbitrary degree distribution $D\sim \{ p_k\}$ on the nonnegative integers. The model is defined as follows (see Figure \ref{fig_config-model} for an illustration). Label the nodes $1, \dots , n$. Draw independent random variables $d_1,\dots ,d_n$ from $D$ and create $d_i$ stubs going out of node $i$, $i=1,\dots ,n$. The network is then constructed by sequentially connecting the stubs pairwise at random. The first stub, of individual 1 say, is connected to any of the remaining stubs uniformly at random. These pair of stubs are then removed, and the procedure is repeated until there are no stubs remaining. This procedure may result in multiple edges between certain pair of nodes, and self loops, and if the number of studs happens to be odd, there will be one remaining stub which cannot be connected. It has been proven that the number of such imperfections are bounded in probability as long as the degree distribution has finite mean $E(D)$, and removing multiple edges (keeping just one), self-loops and one possible odd stub will then have negligible effect on the network and its degree distribution. The configuration network is hence the network obtained after removing multiple edges, self-loops and the possible remaining odd stub.

Another popular network model is the \textbf{Preferential attachment model} due to Barabasi and Albert (1999). This model contains, in its simplest form, one parameter $r$ being a positiv integer, and is defined sequentially starting with a single node without any edges. At each time step $k$ one node equipped with $r$ edges are added to the existing network. The randomness comes from how the $r$ edges are connected to the existing nodes, and this is done by attaching each edge independently, and the probability to attach to a specific node is proportional to its current degree. As a consequence, there is a tendency to attach to nodes already having high degree: the "preferential" feature of the model. This procedure is continued until there are $n$ nodes present in the network (and $r(n-1)$ edges).

The final model we will describe is the \textbf{Small world model} by Watts and Strogatz (1998). In this model all nodes are labelled and put one a line which is made to a circle. The model has two parameters, $k$ and $\rho$, the former being a positive integer and the latter a number between 0 and 1 (typically close to 0 -- sometime it is scaled to $\rho/n$). Each node is first connected to its $k$ closest neighbour on both sides, for example, individual 1 is connected to $2, \dots ,1+k$ and $n,\dots , n-k+1$. Then each edge, independently and with probability $\rho$, rewires one of its end nodes to a uniformly selected node.

The models described above are all for static undirected networks with one type of nodes and no weight on the edges.  There are many other random network models defined for particular purposes. For example Ball et al.\ (2013) define a configuration type model allowing also for arbitrary clustering and degree correlation. 

Other models are dynamic in nature, allowing both individuals to die and get born, and/or for existing edges to disappear and new to appear. Often edges between unconnected pairs of individuals appear randomly in time but at equal rates and existing edges disappear at constant rates, and the population size is either constant equal to $n$ (an individual who dies is replaced by a newborn individual without connections), fluctuates around $n$, or is growing according to a supercritical branching process (e.g.\  \cite{BLT11}). There are also models where the network is affected by the ongoing epidemic, for example individuals distancing themselves from infected individuals (e.g.\ Leung et al., 2018) but these are harder to analyse and will be not discuss further.

If there are different types of individual it is often natural to let the probability of pairs of individuals to be connected to depend on the type of the two individuals. One such model, where the type space is continuous and reflecting "social activity" are the Poissonian random graphs (Norros and Reittu, 2006). There each individual $i$ is given a social activity weight $W_i$ and the probability that individuals $i$ and $j$ are connected is given by $\min (W_iW_j/n, 1)$ or a similar expression. A more specific model is where there are two types of individuals, and where an individual is only connected to individuals of the other type, heterosexual networks and client-server-networks being two examples. Such networks are referred to as bipartite networks (Newman et al., 2002). 

Other random network models allow for directed as well as undirected edges, and edges having different weights, typically reflecting "closeness" of the social relationship (e.g.\ Spricer and Britton, 2015, and Barrat et al., 2004). A different class of network models are \textbf{Random block models}, where individuals can be grouped into a small number of (known or more often unknown) subcommunities, where the probability of two nodes being connected depends on the groups of the two involved individuals (e.g.\ Nowicki and Snijders, 2001). 

Many of the models discussed above can be put under the general framework of inhomogeneous random graph models for which there is a rich theory of results available (e.g.\ \cite{BJR07}).

A slightly different very flexible class of random graph models are socalled \textbf{Exponential random graph models}, ERGMs in short \cite{SPRH06}. This model class is inspired from statistical physics, and allow for penalizing or favouring more or less any network feature. It could feature individual edges, but most often it does so for summary statistics $S_1,\dots ,S_k$, such as mean degree, the number of triangles, high/low degree correlation and higher moments of the degree distribution. Given the set of chosen network statistics and corresponding model parameters $\theta_1,\dots ,\theta_k$, the probability of a specific network/graph outcome $G$ is defined by
$$
P(G)\propto e^{\sum_j\theta_js_j(G)},
$$
where $s_j(G)$ is the value of the summary statistic for the network $G$, an the proportionality constant is given given by the corresponding sum over all $2^{\binom{n}{2}}$ possible networks of size $n$. There is no direct method for generating such networks, instead they are obtained by starting with an initial network and then adding/removing edges in an MCMC like manner until the chain is close to stationarity. Probabilistic properties of such graphs are less explored compared to models discussed above, and they have mainly been useful for inferring the importance of various network measures in small social networks. Their application for epidemics on networks remain mainly yet to be shown.

\section{Infectious disease spreading models}\label{Inf-models}

In the previous section we described several random network models. In the current section we assume the network is given to us, generated from a suitable network model. We now describe some epidemic models for such a network. 

The two models we describe are so called SIR epidemic models, where individuals are first Susceptible, and if they get infected they become Infectious and after a while Recover and become immune. 

We describe first the discrete time epidemic model (Reed-Frost) and then a continuous time Markovian model. Both are defined on a static undirected network/graph $G$.

\begin{defin}[The discrete time Reed-Frost epidemic on a network] Initially, in generation $k=0$, one randomly chosen index case is infectious, and the rest of the population/network is susceptible. Individuals who are infectious in generation $k$, infects each susceptible neighbour in $G$, independently with probability $p$ and then recover (nothing happens with immune or infectious neighbours). Those who become infected by at least one infective become infectious in generation $k+1$. The epidemic goes on until the first generation $T$ when no new infections arise. The number of individuals that get infected during the course of the epidemic (including the index case) is denoted $Z$ and called the "final size".
\end{defin}
\textbf{Remark 3.1}
It is possible to have some other, random or non-random, set of index cases. The model is then the same except that it is started by more than one index case. This model can also be defined in continuous time as described after the next definition.
\vskip.3cm

\begin{defin}[The continuous time Markovian epidemic on a network] Initially, at time $t=0$, one randomly chosen index case is made infectious, and the rest of the population/network is susceptible. While an individual is infectious it has infectious contacts with each susceptible neighbour in $G$ randomly in time according to independent Poisson processes with rate $\beta$. Each infected individual remains infectious for a period $I\sim Exp(\gamma)$ (exponentially distributed with mean $1/\gamma$) efter which it revoers and becomes immune. All infectious periods and contact processes are defined independently. The epidemic goes on until the first time $T$ that there are no infectious individuals and the epidemic stops. The number of individuals that get infected during the course of the epidemic (including the index case) is denoted $Z$ and called the "final size".
\end{defin}

\textbf{Remark 3.2}
It is not hard to show that the Markovian network epidemic model can allow for a random latent period upon infection and before becoming infectious, without affecting who gets infected at the end (but of course affecting its time dynamics). The model can also be extended to let the infectious period $I$ follow an arbitrary random distribution, however then the model is no longer Markovian. One particular choice is when $I\equiv \iota$ (non-random and equal for all individuals), denoted the continuous time Reed-Frost epidemic. It can be shown that the distribution of the final size for this choice of infectious period is identical to that of the discrete time Reed-Frost epidemic if the two models are callibrated by $p=1-e^{-\beta \iota}$. The time dynamics of the two models are however different. 
\vskip.3cm

The two models described above are different in that the first model considers the disease outbreak to occur in discrete time referred to as generations, and the latter in continuous time. In reality epidemic outbreaks of course take place in continuous time. However, if there is a fairly long latency period, followed by a short concentrated infectious period (such as e.g.\ measles and Ebola), then the description of generations makes sense, at least in the beginning of an outbreak. As described in Remark 3.2 there is also a continuous time version of the Reed-Frost network epidemic. The important mathematical difference between the Markovian network epidemic and either of the two Reed-Frost network epidemics lies in that the events of infecting different neighbours are \emph{independent} in the Reed-Frost model whereas they are positively correlated in the Markovian version. This is easy to show mathematically and comes from the fact that, when the infectious period has random length, the event to infect a given neighbour indicates a long infectious period which increases the risk for infecting also other friends. 

In  both models above it is only possible to infect neighbours in the network, where being neighbours reflect a social proximity of relevance for the disease under consideration (e.g.\ daily close contact). For many infectious diseases transmission also occur from more random type of contacts, like sitting next to each other on a bus. The models defined above can add such random contacts as we now define.

\begin{defin}[Network epidemic models with random contacts]
Start with either of the network epidemics defined above (or their extensions described in the remark). The discrete time model is then modified by, at each time step and for each infected, beside infecting susceptible neighbours in $G$ with probability $p$ we now also let them infect, independently, each susceptible in the whole community (neighbour or not) with probability $\beta_G/n$. For the continuous time version where infectives infect each susceptible neighbour at rate $\beta$, we now also let infectives have infectious contacts with each other individual (neighbour or not) independently at rate $\beta_G/n$. 
\end{defin}
\textbf{Remark 3.2} The rate of infecting non-neighbours is hence much smaller than infecting neighbours. The random contact were, for mathematical convenience, defined to happen also with neighbours. The total infection rate with a susceptible neighbour is hence $\beta+\beta_G/n$ which for all practical purposes is equal to $\beta$ when the community is large. The contact rate/probability $\beta_G/n$ with a specific individual outside the household is very small in a large community (as it should be). However, the overall rate that an individual makes random contacts equals $n\beta_G/n=\beta_G$ which is not negligible.
\vskip.3cm

There are numerous extentions to these models. Individuals may be categorized into different types, and the transmission rate/probability can then depend on the two types involved, there may be different types of edges, each type having a specific infection rate/probability, and the network may be dynamic and infection can only take place along currently existing edges.

The models defined above assumed that all individuals beside the index case were susceptible. In reality this may not be the case since there might be prior immunity in the community. If there are immune individuals in the community, such individuals can simply be neglected, and edges connecting to them as well. The degree of susceptible individuals should hence reflect the number of \emph{susceptible} neighbours which should be kept in mind when e.g.\ estimating degree distributions from census data, so-called egocentric network data.

\section{Models for prevention}\label{Prevention}

In the previous section it was mentioned that only initially susceptible individuals should be considered, and that initially immune individuals and their connections can simply be ignored. In the current section we also assume all individuals to be susceptible, but now we see what happens if some are immunized e.g.\ by vaccination (from now on we call them vaccinated). It is then natural to keep track also of the vaccinated individuals in order to study the effect of vaccination. Suppose that a vaccine giving complete immunity is available and that this can be distributed prior to the start of the outbreak. Mathematically, this can be modelled by labelling vaccinated individuals as recovered/immune and to remove them and all their connections from the network. The effect is that the size of the network of susceptibles (unvaccinated) becomes smaller, but more important, that the degrees of remaining individuals are reduced. The practical effect is hence that all vaccinated are protected, but also the non-vaccinated individuals profit in that they have fewer neighbours who can infect them. It is also possible to consider vaccines giving partial immunity (which reduces the risk of getting infected from neighbours) and potentially also reducing infectivity in case of still getting infected (see \cite{HLS10} for such description and how these quantities are estimated) but we will not consider such models further.

If a fraction $v$ of the community are vaccinated prior to the outbreak, it is of interest to see the effect of such a control program as compared to no prevention. In order to study this (which we briefly do inte next section) it is not enough to specify what fraction was vaccinated. Also \emph{who} was vaccinated needs to be known. Needless to say, vaccinating individuals with many neighbours is better from a public health perspective than vaccinating individuals with no or few connections.  

One vaccination policy which could be easy to implement is where candidates are chosen at random, so that the group of vaccinees is a uniformly chosen fraction $v$ of the community. We call this strategy the \emph{uniform vaccination strategy}. There are several other more efficient vaccination strategies, but these are often harder to implement practically. One strategy is to choose the individuals having largest degrees, so the fraction $v$ being vaccinated are those with highest degrees. This is often the best or close to best strategy, but on the other hand it is rarely the case that the degrees of individuals are known. Instead often other proxys are used to reach socially active individuals. For instance, when considering STI's condoms are sometimes distributed freely at night clubs/discoteques and/or at sex-counselling cliniques, thus reaching sexually active people. Another more mathematically formulated strategy is the so called \emph{acquaintance vaccination strategy} \cite{CHB03}. In this strategy individuals are chosen uniformly and then one of its neighbours are vaccinated, and this is done until a community fraction $v$ has been vaccinated. By vaccinating friends of randomly selected individuals, rather than the individuals themselves, individuals being vaccinated will tend to have higher degree. This follows from the somewhat sad network property that "your friends (typically) have more friends than you do". In practice it seems like an unethical vaccination strategy, but a related strategy that has been implemented is to give a vaccine or other protection to randomly selected individuals \emph{and} their partners.

There are other forms of prevention than vaccination and isolation. Mathematically there are two different types of prevention. One aims at reducing the rate of contact between individuals, and the other aims at reducing the risk of transmission upon contact. Isolation belongs to the first group whereas vaccination to the second -- so we have just seen that their mathematical effect may not differ, but often they do. When it comes to preparedness for a new pandemic influenza, school closure is often considered as one option for reducing disease spreading (e.g.\ \cite{CVB08}), but also vaccination once a vaccine has been developed for the new strain (e.g.\ \cite{L05}), and when it comes to more serious diseases spreading locally, like SARS and Ebola, travelling bans are often discussed and their effects modelled (e.g.\ \cite{P14}).

\section{Model properties}\label{Properties}

In the current section we state some results for the network epidemics defined above. We do this without 100\% rigor in order to avoid too many special cases and assumptions. Before doing this we define the most important
quantity of the network epidemic model.
\begin{defin}[Reproduction numbers]
Consider a network epidemic taking place in a large community. The \textbf{basic reproduction number} is denoted $R_0$ and is (loosely) defined as the expected number of new infections caused by typical infected individuals during the early stage of the epidemic. The \textbf{ preventive reproduction number} after a vaccination strategy $S$ with vaccination coverage $v$ has been implemented, is denoted $R_v^{(S)}$. 
\end{defin}
The main reason why these reproduction numbers are important lies in their relation to the threshold value 1 which determines if a major outbreak is possible or not:
\begin{result}
For the network epidemics defined in previous sections (and for a very wide class of epidemic models) the over-all fraction getting infected $\tau_n=Z/n$, where $Z$ is the final number getting infected, satisfies $\tau_n\to 0$ in probability if and only if $R_0\le 1$. Further, a vaccination strategy $S$ with vaccine coverage $v$ results in "herd immunty" (thus surely protecting also unvaccinated) if and only if $R_v^{(S)}\le 1$. 
\end{result}

The consequence of the result is hence that a network epidemic having $R_0>1$ is in danger a large community fraction getting infected (for many models it is also possible to determine this fraction but we omit these type of results here). The aim for any vaccination (or other preventive) program is therefore to reduce the reproduction number to below 1, i.e.\ to obtain $R_v^{(S)}\le 1$. 

It is not alway easy to determine the basic reproduction number for a network epidemic model, and for many networks with complicated structure these are not available. For simpler networks $R_0$ is however known:
\begin{result}
Consider the Erdös-Renyi network, the Configuration model network or the Preferential attachment network, having degree distribution $D\sim \{ p_k\}$ having mean $\mu_D$ and variance $\sigma^2_D$. The basic reproduction number for the Reed-Frost epidemic on these networks is given by
$$
R_0^{(RF)}=p\left( E(\tilde D -1)\right) =p\left( \frac{\sum_kk^2p_k}{\mu_D} - 1\right) = p\left(\mu_D + \frac{\sigma^2_D-\mu_D}{\mu_D}\right).
$$
The basic reproduction number for the Markovian epidemic on these networks is given by
$$
R_0^{(M)}=\frac{\beta}{\beta+\gamma}\left( E(\tilde D -1)\right) = \frac{\beta}{\beta+\gamma} \left( \frac{\sum_kk^2p_k}{\mu_D} - 1\right) = \frac{\beta}{\beta+\gamma} \left(\mu_D + \frac{\sigma^2_D-\mu_D}{\mu_D}\right) . 
$$
The basic reproduction number for the two models also having random contacts is as above but adding the term $\beta_G$ in the Reed-Frost model and the term $\beta_G/\gamma$ in the Markov model. For the preferential attachment model is is known that the degree distribution $D$ has infinite variance, which hence implies that $R_0=+\infty$.
\end{result} 
We will not prove this result, but give a quick heuristic explanation of the first equality -- the latter equalities follow from simple algebra. The degree of infectives during the early stages of an epidemic is different from the degree of individuals in the community at large. Clearly, individuals with degree 0 (i.e.\ having no connections) will never get infected. More precisely, an individual with degree $k$ is $k$ times more likely to get infected than an individual having degree 1. The consequence of this is that the degree of infectives during the early stages of an outbreak equals $k$ has probability proportional to $kp_k$. This so called size-biased degree distribution is denoted $\tilde D$, with outcome probabilities $\tilde p_k=kp_k/\mu_D$. In the beginning of an outbreak all neighbours except the infector will be susceptible, so there are $\tilde D-1$ possible individuals to infect, and the probability to infect any given neighbour is $p$ in the Reed-Frost epidemic and $\beta/(\beta+\gamma)$ in the Markov model. This explains the first equalities above: the expression to the right is the probability of infecting a susceptible neighbour multiplied by the expected number of susceptible neighbours of infected individuals during the early stage of an outbreak. The added term for the models also having random contacts is simply the mean number of such global infectious contacts (all will be with susceptibles during the early stages of the outbreak). 

The corresponding reproduction number after a vaccination program has been initiated is often more complicated to derive. However, the uniform vaccination strategy with vaccination coverage $v$ has a simple form:
\begin{result}
For the epidemic models in the previous result (and more or less all epidemic models), the reproduction number $R_v^{(U)}$ after a fraction $v$, chosen uniformly in the community, have been vaccinated,  relates to the basic reproduction number by the following relation:
$$
R_v^{(U)}=R_0(1-v).
$$
The critical vaccination coverage for the uniform vaccination coverage (the fraction needed to vaccinate to reduce $R_v^{(U)}$ down to 1) is hence given by 
$$
v_c^{(U)}=1-1/R_0.
$$
\end{result}
For vaccination strategies that do better than the uniform strategy, typically by vaccinating individuals with higher degree, the corresponding reproduction number is smaller than that of the uniform strategy with the same coverage. Consequently, the critical vaccination coverage is smaller for such strategies as compared with the uniform. However, other strategies typically do not give explicit expressions for the reproduction number and critical vaccination coverage.

As mentioned earlier we will not prove any of the results mentioned above. The proofs are rather complicated, and several results for probabilistic analysis of network epidemics are still missing. The main reasons for the complications are: random networks are complicated in their own right, to add random epidemic process taking place "on" the network clearly adds complexity, and whether individuals get infected or not depend on the status of neighbours so the events of getting infected are dependent.

\section{Statistical inference}\label{Inference}

We now move to the important area of making statistical inference for epidemics taking place on networks. This is a complicated area which desrves more attention. One reason for the complications lie in the underlying probabilistic complexity described in the previous section. However, a more important reason is the fact that often very little of the epidemic, and the network in particular, is observed. Below we describe a few different data scenarios, and give some important questions deserving attention for each of the data settings. The data could either reflect the final outcome of an epidemic, or disease incidence over time, but also reflect what is known or observed about the underlying network. In Subsection \ref{virus} we also briefly describe inference methodology for outbreaks where also virus sequence data is available, methods which make use of the evolution of the virus by comparing virus sequences of diagnosed cases.

\subsection{Epidemic outbreak on known network}

In some situations, for example diseases spreading through airborne aerosols, like influenza, the most important social structure of relevance for disease spreading is believed to be households. Since this is a known structure which is easy to collect information about,  a model for the underlying network is not needed. Consider for example the final outbreak taking place in a community built up of households. That is, the data we observe is $\{ n_{h,i}; 0\le i\le h\le h_{\max} \}$, where $n_{h,i}$ denotes the number of households having $h$ initially susceptible and in which $i$ of them were infected during the outbreak. As mentioned earlier, it is important to only consider susceptible individuals. In practice this can be achieved by testing for antibodies prior to the outbreak.

We have not described a household epidemic model, but the simplest, Reed-Frost type, household epidemic is given by the Reed-Frost epidemic model with random contacts defined earlier, but where the network consists of small, fully connected subgroups -- the households. 

The model can be approximated by assuming that households behave independently and all individuals having a fixed probability $p_G$ to be infected from outside the household. For this approximate model it is possible to write down the probability that $i$ out of $h$ individuals in a household get infected $\pi_{h,i}=\pi_{h,i}(p, p_G)$, as a function of the two model parameters $p$ and $p_G$. This is done by first conditioning on how many that get infected from outside the household (a binomial outcome) and then the rest being infected from household members. These latter probabilities are non-trivial and given by certain recursive formulae not presented here (see Longini and Koopman \cite{LK82} and Addy et al.\ \cite{ALH91}  who also considers different types of individual). The conclusion is anyway that the likelihood is simple in terms of these probabilities:
$$
L(p,p_G)=\prod_{h=1}^{h_{\max}}\prod_{i=0}^h \pi_{h,i}(p, p_G)^{n_{h,i}}.
$$
It is worth emphasizing that this is not the exact likelihood for the stochastic household model, but an approximation relying on a large community where households are close to independent, so it can be called a pseudo-likelihood. Parameter estimates are obtained by maximizing the likelihood with respect to $p$ and $p_G$ (or when the original parametrization with individual global transmission probability $\beta_G/n$ is used then $p_G$ should be replaced by the community infection probability $1-e^{-\beta_G\tau}$, $\tau$ being the overall fraction infected). For details on this we refer to Longini and Koopman \cite{LK82}. The pseudo-likelihood gives a consistent estimate of the parameters $p$ and $p_G$ but their uncertainty estimates are biased from neglecting depedencies -- in Ball et al.\ \cite{bms97} it is described how to correct for this using the proper stochastic epidemic model and its central limit theorem.

\begin{figure*}[h]
\centering
\includegraphics[scale=0.6]{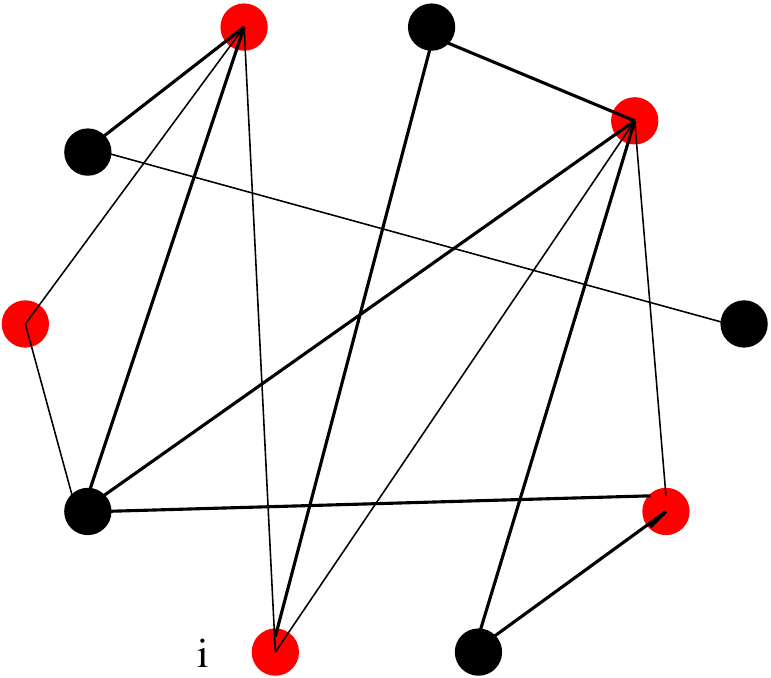} 
\caption{Illustration of the final outcome in a small random social network. Red nodes have been infected and black nodes have not.}\label{fig_network-outbreak}
\end{figure*}

If we instead consider an arbitrary but known network where we observe the final outbreak as illustrated on the small network in Figure \ref{fig_network-outbreak}, inference is much less straightforward even when all transmission takes place along edges in the network and random infectious contacts are not considered. 
The main reason is that we do not observe who-infected whom, or not even in which order individuals were infected. As an illustration, individual $i$ in the figure was infected during the outbreak and so were 2 out of 3 neighbours of $i$. What is the contribution of this to the likelihood? The answer will depend on when $i$ was infected in relation to the neighbours. Clearly $i$ was infected by one of the two infected neighbours (excluding the small probability that $i$ was the index case), but was $i$ infected by the first of the other two getting infected and did $i$ then infect the other, or was $i$ not infected by the first of them and failed to infect any of the remaining individuals? The former would indicate a larger value of $p$ compared to the latter, but information on which of these scenarios (or other scenarios) is not contained in the data thus making inference much harder. One possible approach is to try to collect also information on time of infection (or diagnosis) making the likelihood manageable, or to augment the final size data with this information and use MCMC methods thus integrating over possible and likely temporal outbreak data (see e.g. Britton and O'Neill \cite{bo02}). Another observation from the figure is that there is a higher tendency for individuals with high degree to get infected. There are however exceptions: the individual left to $i$ has four neighbours, and all of them were infected, but still this individual escaped infection by chance.

It seems like an important open problem to come up with simpler pseudo-likelihood methods for the type of data considered. It is not clear if the likelihood is better to consider for different \emph{nodes} or for different \emph{edges} in the network. If considering the likelihood contribution from different individuals, such methods might make use of the fact that the likelihood contribution from an individual $i$ escaping infection should be $(1-p)^{T_i}$, where $T_i$ denotes the total number of infected neighbours to $i$. If instead considering the likelhoood in terms of edges, then edges where both individuals escape infection there is no contribution to the likelihood, and for edges where one individual is infected and the other escapes infection the likelihood contribution is $1-p$, since there was no transmission through that edge. The problem lies in edges where both individuals were infected. For such edges it is not (always) clear if a transmission took place or if the second individual to get infected escapes infection from that neighbour and only later got infected from another neighbour.

It is also possible to consider other known network structures, such as schools, together with households and possible random contacts. The likelihood of course becomes more complicated and some computer intensive statistical methodology has to be used (e.g.\ \cite{CVB08}). 

If temporal data is available inference is often simplified, at least when the order of infection can be inferred from the temporal data. Suppose for simplicity that we observe when individuals get infected and when they recover. We summarize this data by the following quantities for each individual $i$: $(E_i, N_i, I_i)$, where $E_i$ is the accumulated exposure to infection up until individual $i$ was infected, or to the end of the epidemic if $i$ was not infected. $N_i$ and $I_i$ are only relevant if $i$ got infected. In that case $N_i$ denotes the number of infectious neighbours of $i$ at the time when $i$ gets infected, and $I_i$ denotes the length of the infectious period of $i$. Clearly, this information is available when the network is known and we observe infection and recovery times of all infected individuals. The likelihood for this data is given by:
$$
L(\beta, \gamma )=\prod_ie^{-\beta E_i}\prod_{i\in Inf} (\beta N_i) \prod_{i\in Inf} f(I_i) \propto e^{-\beta\sum_iE_i} \beta^{Z} \prod_{i\in Inf} f(I_i),
$$
where we have left out a combinatorial factor not affecting estimation on the left hand side, $Z$ denotes the final number infected, and $f(I_i)$ is the density of the infectious period -- this likelihood is valid also for other distributions than the exponential distribution. From this we see that estimates of $\beta$ and $\gamma$ are more or less independent, and the inference for the infectious period distribution is just as when observing i.i.d.\ infectious periods. Further, it is easy to show that the ML-estimate for $\beta$ is given by 
$$
\hat\beta =Z/\sum_iE_i,
$$
the number of infections dived by the overall exposure times between infectious individuals and their susceptible neighbours. To obtain standard errors for this estimate remains an open problem for most networks. However, using theory for counting processes (Andersen et al., \cite{abgk93}) this should be possible to derive using similar to techniques as for homogeneous epidemic models observed continuously in time (see for example Diekmann et al.\ \cite{DHB13}, Section 5.4.2).

\subsection{Epidemic outbreak on known network model}

Much more common than knowing the underlying social network (as considered in the previous subsection) is to not know the underlying network. In some situations the \emph{structure} of the underlying network could be known, at least partially, but it is not known at an individual level. This would be the case when a (representative) sample of so-called egocentric networks are observed but not related to the epidemic outbreak  (e.g.\ \cite{PHLH11}). As an illustration for STI's in a one-sex population, a study of sexual habits can be used to infer the degree distribution $\{p_k\}$ for the number of sex-partners the last year, and then the configuration model or the preferential attachment model (for suitable choice of $r$) might be used to model the underlying sexual network even though it is not observed. 

If an epidemic outbreak is observed for such a network, the inference methodology will depend on what is observed. If all that is observed is the final fraction getting infected, then very little can be done. If the underlying network model, including parameter values is known, then it is in principle possible to estimate one disease spreading parameter, at least for network epidemics for which the limiting final size has been derived. This final size limit $\tau$ will be a function of the network model (assumed known) and disease model parameters (e.g.\ $p$ in the Reed-Frost version), and the estimate $\hat p$ is then the value of $p$ for which $\tau$ coincides with the observed final fraction $\tilde \tau$. Such an estimate of disease spreading paramaters relies on the network model, including its parameters, is fully known. If for example an incorrect value for the mean degree $\mu$ is used, then $\hat p$ will be biased.

If instead incidence is observed over time, and/or also network information of infected individuals is collected, then inference procedures should be possible to improve, but such methods are still to be developed.

\subsection{Epidemic outbreak on unknown network model using virus sequences}\label{virus}

A related situation as considered in the previous section is for epidemic data where diagnosed individuals are also sequenced, meaning that the DNA of the virus (or other disease agent) is sequenced. The underlying idea is that, for viruses that evolve (within infected individuals) at the same time scale as the epidemic spreads, individuals close to each other in the transmission tree will have virus sequences that are more similar as compared to individuals that are farther away from each other in the transmission tree. So by observing also the sequences among diagnosed individuals, it is possible to learn more about the underlying transmission tree and hence about the underlying social network upon which the disease spreads.

This new research area, of making inference using sequences, with or without traditional epidemic data, has exploded the last 10-15 years. To describe this field in detail would require many pages of modelling and statistical methodology -- here we simple sketch it briefly. For a more thorough descriptin we refer to other papers, for example the survey chapter by Klinkenberg, Colijn and Didelot in Held et al. \cite{HHNW19}.

\begin{figure*}[h]
\centering
\includegraphics[scale=1.0]{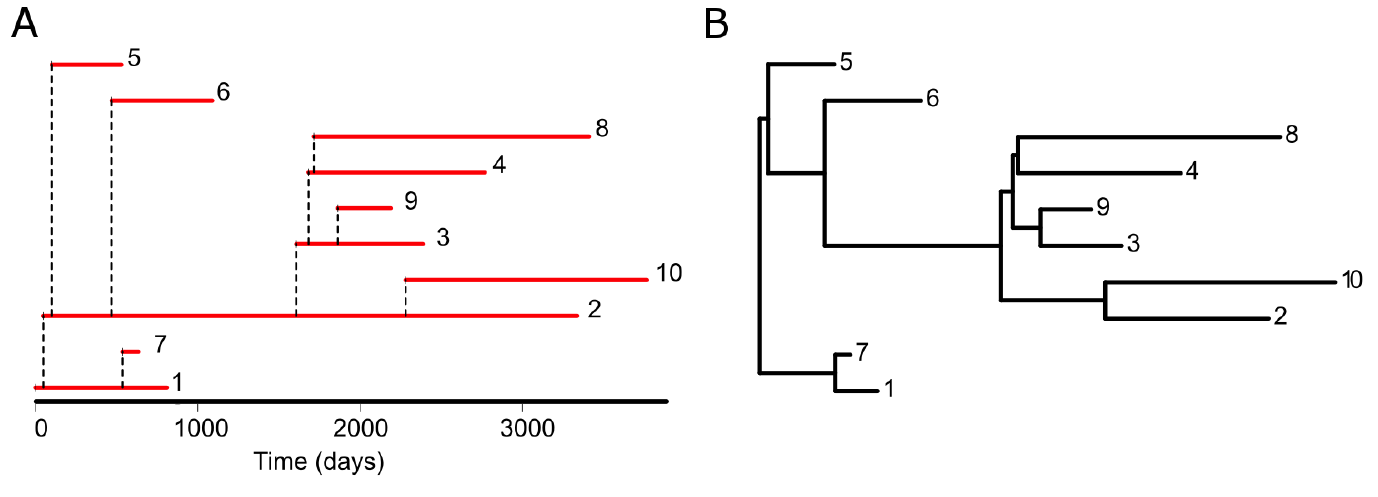} 
\caption{Illustration of an transmission tree (A) and the corresponding virus genealogy (B). In the transmission tree individual 1 is the index case and this individual directly infects individual 2, and later individual 7 (infections are indicated by dashed upward lines, and infectious periods of individuals are marked with red horizontal lines). In the virus genealogy the information about who infected whom is lost, as are the starting times of individuals' infectious periods. The distance unit for the transmission tree is calendar time whereas it is evolutionary distance in the virus genealogy. }\label{Infectiontree}
\end{figure*}

There are numerous models for how a virus population within a host evolves over time. A simple model assumes that an individual is infected by one single virus strain, and that this strain evolves over time within the individual, but that all copies of the virus are identical. The motivation is that mutations that are inferior quickly die out whereas  mutations that are superior quickly take over the entire virus population -- this is called the ``one dominant strain'' assumption as opposed to allowing for within-host diversity. When applying the one-dominant-strain assumption and a suitable evolutionary model for DNA mutations (see e.g.\ Felsenstein \cite{F03}) it is possible to infer the virus genealogy of the virus sequences from the infected individuals. This virus genealogy, where distance is measured in evolutionary distance (e.g.\ number of mutations per 1000 base pairs), is in turn related to corresponding transmission tree where distance is measured in calender time. In case of one-dominant-strain and assuming a constant molecular clock (i.e.\ constant rate of mutation over time and across individuals) the two trees are identical, except that the virus genealogy does not contain information on who infected whom, which the transmission tree does (cf.\ Figure \ref{Infectiontree}) for an illustration).

So by sequencing viruses of infected individual, it is possible to learn about the virus genealogy and hence also about the transmission tree. By comparing the inferred virus genealogy with typical virus genealogies from various network and transmission models (and their parameters) it is possible learn what the underlying network structure might have been. These type of ideas are often used for outbreaks of HIV in order to learn more about spreading patterns (e.g.\ Leventhal et al.\ \cite{L12} and Giardina et al.\ \cite{G17}).

The statistical methodology is often quite involved and numerically intensive, for example employing MCMC, Approximate Bayesian Computation (ABC) and Iterated filtering methods. As mentioned earlied the main idea in MCMC for infectious disease data is to treat unknown features, such as the underlying network, as latent variables used in the MCMC chain which are then integrated over. The main idea with ABC and iterated filtering methods is to simulate/generate output for different choices of models and parameters, and to run additional simulations for models/parameters ''close'' to those of earlier simulations which resembled the observed data (importance sampling). The comparison of how well a simulation agrees with data can for example be performed using tree shape measures such as Sackin index (cf.\ \cite{L12}).

A disadvantage with most work in this new area of infectious disease inference is that traditional epidemic data, in particular of exposed individuals avoiding infection, is rarely used. In fact, Li et al.\ \cite{LGF17} even report that precision in inference worsens when the statistical analysis also make use of incidence data beside the virus sequence data. It is my strong opinion that this should not be the case if a correctly performed statistical analysis is used on a suitable statistical model for community structure and disease transmission. It is an important area to develop statistical methodology for network epidemic data using both virus sequences and incidence and other epidemic and network data.

\subsection{Predicting effects of preventive measures}

One of the main reasons for mathematical modelling and statistical analysis of (network) epidemics is prevention: next time there is a similar epidemic outbreak, or even during an ongoing outbreak, it is of interest to predict what would happen if various control measures are put in place. These control measures could for example be vaccination, increased condom use, isolation of infected cases, school closures or introducing travelling restrictions. The common way to proceed is to estimate paramaters for a suitable model of the network and disease spreading, and then to mathematically analyse what would be the outcome if some control measure was put in place for the studied model with parameter values taken from the estimation step. As a very simple illustration, suppose the basic reproduction has been estimated to $\hat R_0=1.5$ and with standard error $s.e.(\hat R_0)=0.1$. If a vaccine giving 100\% immunity is available next time and a fraction $v=0.2$ of randomly selected individuals were vaccinated, an estimate of the new reproduction number would be $\hat R_v^{(U)}=\hat R_0(1-v)= 1.5*0.8=1.2$ (cf.\ Result 5.3 in Section 5). Similarly, the critical vaccination coverage is estimated to $\hat v_c^{(U)}=1-1/\hat R_0=0.33$, meaning that predicted fraction necessary to vaccine in order to avoid a future outbreak is 33\%. Using also the standard error it is possible to construct an upper confidence bound on $v_c^{(U)}$.

In more complicated situations, like an ongoing epidemic where vaccination is introduced or modelling effects of school closure, in structured communities, the corresponding effects are most often studied by means of simulations (e.g.\ Longini et al.\ \cite{L05} and Cauchemez et al.\ \cite{CVB08}). The basic idea is however the same: to first use data to infer model parameters and then to study effects of intervention for the model and its estimated parameters. The effect of intervention, for example how much susceptibility is reduced by the vaccine, or the effect of closing schools on infection between school children, has to be known or estimated using some other data source.

Increasing knowledge in this area can either consist of improving the mathematical analysis of effects of intervention in network models, and/or to improve inference procedures for data on network epidemics.

\section{Discussion and Extensions}

The ambition with the paper has been to give some basic ideas behind stochastic modelling and statistical analysis of infectious disease outbreaks taking place in communities structured as social networks. The focus has been on methodology rather than hands-on data analysis. This area uses many different probabilistic and stochastic techniques making it impossible to cover the area in any detail -- instead we have referred to useful publications in different subareas where more can be learned.

An inherent problem with the area, beside its mathematical complexity stemming from randomness in both the underlying network and the disease spreading on it, is that very rarely is the network structure even partly known or observed. The area certainly requires more research, in particular to make better use of virus sequences and other proxies (e.g.\ from egocentric networks) giving information about the underlying network, and to combine such information with incidence and exposure data for an improved combined statistical analyis.

Needless to say, we have left out several important issues. Some very important issues only briefly mentioned in the text, and important to consider both in modelling and statistical analysis are: under-reporting, asymptomatic cases and partial immunity. Nearly all data on epidemic outbreaks miss some infected cases, perhaps because they were asymptomatic or mild cases, but also for other reasons. Partial immunity is another aspect which cannot be neglected. If estimation is performed in a partially immune community, then the conclusion are no longer valid when immunty in the community wames.  

We now mention some topics we have not touched upon: models and analysis of diseases that are endemic in the population, models and analysis for situations where indivuduals change behavior over time -- perhaps as a result of the epidemic outbreak, statistical analysis of epidemics on dynamic networks, models for which the infectivity of an individual varies over time (e.g.\ the accute and chronic phase of HIV), and the important area of model fit.

As has been mentioned in several places, the statistical methods for models capturing both (often unobserved) network structure and transmission models, are often too complicated for direct methods such as maximum likelihood estimation. Instead some numerically intensive method like MCMC, ABC or Particle filtering can be adopted. Held et al.\ \cite{HHNW19} is a recent book describing such (and other) statistical methods for infectious disease data, but without focus on the network situation. 

Eventhough statistical analysis of network epidemics is quite hard, it is encouraging to see that preventive measure make use of ideas from network epidemic modelling. For example, distribution of vaccines and/or condoms to selected individuals and their \emph{partners} (mimicking the acquaintance vaccination strategy in Section 4), and ring-vaccination in Ebola outbreaks, clearly making use of social structures by vaccinating relatives and friends of all reported cases \cite{HCL17}.

It is my strong belief that much progress can be made in this important research area by development of new statistical methodolology together with close connections to data collectors, thus paving the way to more informative data collection as well.

\section*{Acknowledgements}

 I am grateful to the Swedish Research Council (grant 2015-05015) for financial support.

\end{document}